# Evidence for anisotropic spin-triplet Andreev reflection at the 2D van der Waals ferromagnet/superconductor interface

Ranran Cai[1,2][†], Yunyan Yao[1,2][†], Peng Lv[3], Yang Ma[1,2], Wenyu Xing[1,2], Boning Li[1,2], Yuan Ji[1,2], Huibin Zhou[1,2], Chenghao Shen[4], Shuang Jia[1,2,5,6], X. C. Xie[1,2,5,6], Igor Žutić[4], Qing-Feng Sun[1,2,5,6] & Wei Han[1,2]*

[1]International Center for Quantum Materials, School of Physics, Peking University, Beijing 100871, P. R. China
[2]Collaborative Innovation Center of Quantum Matter, Beijing 100871, P. R. China
[3]Department of Physics, Wuhan University of Technology, Wuhan 430070, China
[4]Department of Physics, University at Buffalo, State University of New York, Buffalo, New York 14260, USA
[5]CAS Center for Excellence in Topological Quantum Computation, University of Chinese Academy of Sciences, Beijing 100190, P. R. China
[6]Beijing Academy of Quantum Information Sciences, Beijing 100193, P. R. China
[†]These authors contributed equally to the work
*Correspondence to: weihan@pku.edu.cn

**Fundamental symmetry breaking and relativistic spin–orbit coupling give rise to fascinating phenomena in quantum materials. Of particular interest are the interfaces between ferromagnets and common s-wave superconductors, where the emergent spin-orbit fields support elusive spin-triplet superconductivity, crucial for superconducting spintronics and topologically-protected Majorana bound states. Here, we report the observation of large magnetoresistances at the interface between a quasi-two-dimensional van der Waals ferromagnet $Fe_{0.29}TaS_2$ and a conventional *s*-wave superconductor NbN, which provides the**

**possible experimental evidence for the spin triplet Andreev reflection and induced spin-triplet superconductivity at ferromagnet/superconductor interface arising from Rashba spin-orbit coupling. The temperature, voltage, and interfacial barrier dependences of the magnetoresistance further support the induced spin-triplet superconductivity and spin-triplet Andreev reflection. This discovery, together with the impressive advances in two-dimensional van der Waals ferromagnets, opens an important opportunity to design and probe superconducting interfaces with exotic properties.**

Fundamental symmetry breaking and relativistic spin–orbit coupling give rise to interesting phenomena in quantum materials[1,2]. For over sixty years, the interplay between ferromagnetism and superconductivity, has offered a wealth of intriguing phenomena in ferromagnet (FM)/superconductor (SC) heterostructures[3-6]. However, to overcome a strong suppression of spin-singlet superconductivity by the FM's exchange field the platforms supporting spin-triplet pairing are sought. They are desirable for dissipationless spin currents in superconducting spintronics[5-7], and probing quantum materials[8], as well as for realizing elusive Majorana bound states to implement topological quantum computing[9,10]. The common expectation that spin-triplet pairing in superconducting spintronics requires complex FM multilayers, typically relying on noncollinear/spiral magnetization or half metals[3-6,11].

Here, we report the possible experimental evidence for the spin-triplet Andreev reflection and induced spin-triplet superconductivity at the interface of a quasi-2D van der Waals (vdW) FM and a conventional $s$-wave SC with Rashba spin-orbit coupling (SOC). Such vdW heterostructures offer a great versatility in exploring the interplay between ferromagnetism and superconductivity, beyond the lattice-matching constraints of all-epitaxial FM/SC heterostructures[12]. Our results pave the way for future studies on spin-triplet superconductivity[13,14] and the formation on Majorana bound states[9,10], as well as many normal-state spintronic applications[15].

In contrast to the conventional Andreev reflection at the FM/SC interface (Fig. 1a), an incident spin-up electron forms a spin-singlet Cooper pair in the ordinary SC with a reflected spin-down hole in the FM, spin-triplet Andreev reflection generates the spin-up hole with an injection of an equal-spin triplet Cooper pair in the spin-triplet SC (Fig. 1b). Due to Rashba SOC[16,17], spin-rotation symmetry is broken for the superconducting pairing (Fig. 1c), which acts as a spin-mixing described in conventional FM/SC heterostructures[5,6]. The broken spin-rotation symmetry leads to the spin-singlet paring (m = 0, $S$ = 0) ($S$ is the total spin quantum number, and m is magnetic quantum number) with an unpolarized spin-triplet component (m = 0, $S$ = 1)[18]. The spin-triplet component results in the interface spin-triplet Andreev reflection at the FM/SC interface which is highly anisotropic (Supplementary Note 1 and Supplementary Fig.1), depending on the relative orientation between the magnetization (**M**) in the FM and the interfacial spin-orbit field[14,19]. **M** sets the spin-quantization axis, and unpolarized spin-triplet component (m = 0, $S$ = 1) is projected onto the spin-quantization axis to generate the equal-spin-triplet component (m = 1, $S$ = 1), which can be considered as a spin-rotation process[5]. For example, for FM magnetization along z axis (perpendicular to the interface), unpolarized spin-triplet Cooper pairs component ($|S = 1, S_y = 0\rangle$ and $|S = 1, S_x = 0\rangle$ can be projected to the spin quantization axis as $|S = 1, S_z = 1\rangle$ due to the spin rotation process ($S_y$, $S_x$, and $S_z$ are the spin quantum numbers along *y*, *x*, and *z* direction, respectively). Thus, both the $|S = 1, S_y = 0\rangle$ and $|S = 1, S_x = 0\rangle$ components will contribute to the interface conductance when **M** is perpendicular to interface, as illustrated in Fig. 1d. On the other hand, when the **M** is parallel to the interface along *y* direction, the equal-spin triplet Cooper pairs ($|S = 1, S_y = 1\rangle$) can only be projected from unpolarized spin-triplet pairing component $|S = 1, S_x = 0\rangle$, since $[S_x, S_y] \neq 0$. Consequently, spin-triplet Andreev reflection conductance channel is suppressed when **M** is parallel to interface, as illustrated in Fig. 1e. As a result of the anisotropic spin-triplet Andreev reflection processes, there is a low- (high-) resistance state for **M** out-of-plane (in-plane) (Figs. 1d and 1e). Hence, the spin-triplet Andreev reflection can lead to the tunnelling anisotropic magnetoresistance (MR) at the FM/SC interface, a proposed hallmark of the interfacial SOC and spin-triplet superconductivity in FM/SC heterostructures[14,19].

To experimentally probe the anisotropic spin-triplet Andreev reflection and spin-triplet MR, we fabricate the FM/SC devices (see Methods for details), which consist of a quasi-2D vdW $Fe_{0.29}TaS_2$ flake, several *s*-wave superconducting NbN electrodes, and two normal metal Pt electrodes (Fig. 2a and Supplementary Fig. 2). At the interface between the quasi-2D vdW $Fe_{0.29}TaS_2$ flake and *s*-wave NbN electrode, the Cooper pairing consists of both spin-singlet (m = 0, S = 0) and spin-triplet components (m = 0, S = 1) due to the spin-rotation symmetry breaking by the interfacial Rashba SOC (right panel of Fig. 2a). The superconducting critical temperature of the NbN electrode is $T_{SC} \sim 12.5$ K (Supplementary Fig. 3a) characterized by standard four-probe electrical measurement. $Fe_{0.29}TaS_2$ flakes are typical quasi-2D vdW FM, with a Curie temperature, $T_{Curie}, \sim 90$ K, characterized by anomalous Hall effect (Supplementary Fig. 4)[20]. The magnetic easy axis is perpendicular to the sample plane, and **M** of $Fe_{0.29}TaS_2$ can be controlled by a large external magnetic field (**B**) (Supplementary Note 2 and Supplementary Fig. 5). For an in-plane **B** = 9 T, **M** is almost in plane, 83° from the *z* direction. Under **B** = 9 T, the current-voltage characteristics of the NbN electrode are measured, with critical currents of ~ 50 μA at $T = 2$ K (Supplementary Fig. 4b). Typical *dI/dV* curves of the $Fe_{0.29}TaS_2$/SC junctions as a function of *T* and *B* are shown in Supplementary Note 3 and Supplementary Fig. 6.

To characterize the expected MR arising from anisotropic spin-triplet Andreev reflection, the interfacial resistance between the quasi-2D vdW FM $Fe_{0.29}TaS_2$ and SC electrode is measured using the three-terminal geometry (Fig. 2a and Methods). Figure 2b shows the typical MR curve (blue) measured (device A; Supplementary Fig. 2) as a function of the magnetic field angle in the $yz$ plane ($\Theta_{yz}$) at $T = 2$ K and $\mathbf{B} = 9$ T. The observed MR shows a strong correlation with $B$-controlled $\mathbf{M}$ (Supplementary Fig. 7). In contrast to this large MR at $T = 2$ K, the normal-state interfacial resistance exhibits little variation at $T = 20$ K. A possible important contribution of vortices in type-II SC to the observed MR has been ruled out from our control measurements on normal metal/SC heterostructures at $T = 2$ K (orange curve in Fig. 2b; Supplementary Note 4 and Supplementary Fig. 8). We have also fabricated the control devices of $Fe_{0.29}TaS_2/Al_2O_3$/normal metal (Al), where no MR could be observed at $T = 2$ K (Supplementary Fig. 9), which further indicates the critical role of SC for the observed MR. Furthermore, the $\pi$-periodic oscillation further supports that the observed MR results from the anisotropic feature of spin-triplet Andreev reflection at the interface with Rashba SOC[14]. Figure 2c shows the MR results measured on the device B as a function of $\Theta_{yz}$ at $T = 2$ K and $\mathbf{B} = 9$ T. The MR ratio can be defined as:

$$\mathrm{MR}(\Theta_{yz}) = \frac{R(\Theta_{yz}) - R(\Theta_{yz} = 0)}{R(\Theta_{yz} = 0)} \times 100\% \qquad (1)$$

The $R(\Theta_{yz} = 0)$ and $R(\Theta_{yz} = 90)$ are the interfacial resistances for magnetic field that is perpendicular and parallel (along $z$ and $y$ directions in Fig. 2a) to the FM/SC interface, respectively. Interestingly, the observed MR ratio is ~ 37% ± 2% for device A, and ~ 103% ± 4% for device B, which are much larger than previous reports on the tunneling anisotropic MR in FM/semiconductor heterostructures arising from the Rashba and Dresselhaus SOC[16,17].

Next, we investigate the temperature evolution of the MR to distinguish the contributions from the spin-triplet Andreev reflection and spin-dependent scattering by Bogoliubov quasiparticles under large magnetic field. Figure 3a shows the MR ($\Theta_{yz}$) for device B at $T$ = 2, 4, 8, and 9 K, respectively, under the magnetic field of $B$ = 5 T. Figure 3b summarizes temperature dependence of the MR ratio for device B measured at **B** = 9, 7, and 5 T, respectively. The MR appears for $T < T_C$, and starts to saturate below the temperature of ~ 5 K. The MR is no longer observable for $T \sim T_C$ at **B** = 9, 7, and 5 T (Supplementary Fig. 10). Clearly, there is no enhancement or any anomaly of the MR observed at the temperature slightly below $T_C$, which further confirms that contribution from spin-dependent scattering by Bogoliubov quasiparticles is negligible[21,22].

To further investigate the MR at the quasi-2D vdW FM $Fe_{0.29}TaS_2$/SC interface, we systematically vary the bias voltage ($V_{bias}$), which also affect the junction voltage ($V_{3T}$) across the interface. At the interface, the induced SC energy gap ($\Delta_{In}$) by SC proximity effect with spin-triplet component is smaller compared to the SC gap ($\Delta_{NbN}$) of bulk NbN electrode, as illustrated in Fig. 4a. When the potential ($eV_{3T}$) of the incoming electrons is considerably smaller than the interface spin-triplet superconducting energy gap ($\Delta_{In}$) (Fig. 4a), the charge transport channel is dominated by the anisotropic spin-triplet Andreev reflection. Hence, the spin-triplet MR exhibits little variation with the $eV_{3T}$ within the $\Delta_{In}$. As the $V_{3T}$ increases, other isotropic transport processes, such as electron-like and hole-like tunneling transmissions[14], also contribute to the interface conductance. As these transport processes are **M**-independent, the spin-triplet MR ratio is expected to decrease significantly. Since the change of $V_{3T}$ is much smaller than $V_{bias}$ during the rotation of the external magnetic field, the junction voltage for $\Theta_{yz} = 0$ ($V_{3T\_0}$) is used to qualitatively show the interface voltage dependence of the spin-triplet MR. Figures 4b and 4c summarize these results measured on devices B and C. For small $V_{3T\_0}$, the MR exhibit little variation as the voltage changes. However, when $V_{3T\_0}$ is higher than a critical value, MR strongly decreases as $V_{3T}$ increases. The critical junction voltage is obtained to be ~ 0.15 mV (~ 0.2 mV) for device B (C). We note that at 2 K the thermal energy is $k_B T$ ~ 0.17 meV, comparable to the critical electron potential from the bias-dependent results. Therefore, an accurate value of the proximity-induced superconducting gap is not able to be clearly resolved here, which will need future studies. Additionally, the bias dependence of the spin-triplet MR further confirms that the observed MR is correlated to the sub-gap properties,, and is completely different form *B*-induced spin-splitting density of states at the gap edges of SC electrodes[23].

As the spin-triplet Andreev reflection depends strongly on the FM and SC wave-function overlap, it is expected that the dimensionless interface barrier strength (*Z*) plays an important role in the spin-triplet MR[24,25]. To explore the influence of interface barrier strength on the observed spin-triplet MR, we investigate more than dozen devices that are fabricated with $Al_2O_3$ layer of different thickness (~ 1 - 2.5 nm) between the quasi-2D vdW FM $Fe_{0.29}TaS_2$ and NbN SC electrodes. This process leads to a large range of interface resistance area product ($R_J S$) from ~ 10 to ~ 2000 Ω μm$^2$, resulting in the FM/SC heterostructures with very different *Z*-values. Figure 5 shows the measured MR ratio as a function of the $R_J S$ at $T$ = 2 K and $B$ = 9 T (Note: the MR is not observable for very large $R_J S$ and not plotted in this figure). The largest MR is observed with $R_J S$ ~ 48.4 Ω μm$^2$. The strong correlation of the MR ratio and $R_J S$ reveals the important role of the *Z*-value in the spin-triplet MR.

This surprising nonmonotonic MR dependence on $R_J S$ agrees well with the theoretical expectations[14,25]. The effective barrier strength is modified by SOC and depends on the helicity (outer/inner Rashba bands, Fig. 1b), $Z_\pm = Z \pm \bar{\gamma} k_\parallel$, where $\bar{\gamma}$ is the SOC parameter[25] and $k_\parallel$ is the component of the wave vector along the interface (Fig. 5 inset). At zero $k_\parallel$, the vanishing of Rashba

SOC does not support spin-triplet component. At nonzero $k_\parallel$, increasing $Z$ can reduce $|Z_+|$ or $|Z_-|$ and thus enhance such a transmission for a given helicity. For much larger $Z$, all of the conduction channels, including spin-triplet Andreev reflection, are suppressed due to the low interface transparency. As a result, the spin-triplet Andreev reflection and spin-triplet MR will also be nonmonotonic in $Z$. Taken together, the observed nonmonotonic MR dependence with $R_JS$ (Fig. 5) and MR decrease with $T$ or an applied voltage (Figs. 2-4) are all possible experimental evidence for the spin-triplet Andreev reflection in our vdW heterostructures. We note that the spin-triplet MR theory is developed using an idealized model of ballistic systems[14, 25], the role of disorder, which could induce reflectionless tunneling, is expected to reduce the MR amplitude. To the best of our understanding, the spin-triplet Andreev reflection is the major cause for the observation of large MR up to ~103% $\pm$ 4%, and can qualitatively explain the bias and temperature dependence of the MR. Given the growing interest in systems that could support spin-triplet superconductivity, in the future studies, it would be important to generalize our description and also include the effects of disorder and diffusive transport on spin-triplet MR.

Our experimental obervation of a large tunneling anisotropic MR in quasi-2D vdW FM/s-wave SC heterostructures up to ~103% $\pm$ 4% is already promising for spintronic applications and much larger than for the normal-state transport in previously measured heterostructures with a single FM layer[16,17]. More importantly, this result also reveals an emergent spin-triplet superconductivity which, through spin-triplet Andreev reflection, is a sensitive probe of interfacial Rashba SOC. With the advances towards high-quality vdW heterostructures, we anticipate that the magnitude of such spin-triplet MR can be further enhanced and strongly modulated using different 2D vdW FMs due to their highly-tunable Rashba SOC by electric fields[26-29]. This tantalizing opportunity to implement FM/SC heterostructures to design and probe interfacial SOC offers an important boost for superconducting spintronics[3-6,30,31] and Majorana bounds states[9,32]. Furthermore, our quasi-2D platform of proximity-induced spin-triplet superconductivity, combined with the gate-controlled 2D vdW ferromagnetism[28,29,33] could provide tunable magnetic textures to create synthetic SOC[34] and braid Majorana bound states[35].

**Methods**

**Device fabrication**

The quasi-2D vdW $Fe_{0.29}TaS_2$/SC spin-triplet MR devices were fabricated as follows. First, bulk single crystalline $Fe_{0.29}TaS_2$ were grown by the iodine vapor transport method. Then the quasi-2D vdW $Fe_{0.29}TaS_2$ flakes were mechanical exfoliated from the bulk single crystal onto the $SiO_2$ (~ 300 nm)/Si substrates[18]. Second, a first-step electron-beam lithography was used to define the SC electrodes on the quasi-2D vdW $Fe_{0.29}TaS_2$ flakes. The SC electrodes consist of ~ 5 nm thick Nb and ~ 60 nm thick NbN, which were grown in a DC magneton sputtering system with a base pressure of ~ $1.2 \times 10^{-4}$ Pa. Prior to the growth of SC electrodes, a thin $Al_2O_3$ layer (~ 1 - 2.5 nm) is deposited as the barrier to tune the interface coupling strength between the quasi-2D vdW $Fe_{0.29}TaS_2$ flakes and the SC electrodes. The $Al_2O_3$ layer was grown by DC magnetron sputtering with Al target under the oxygen atmosphere. Then, a second-step electron-beam lithography was used to define the two normal Pt electrodes (~ 80 nm) on the quasi-2D vdW $Fe_{0.29}TaS_2$ flakes. The Pt electrodes were deposited by RF magneton sputtering in a system with a base pressure lower than $6.5 \times 10^{-4}$ Pa. The optical images of three typical devices (A, B, and C) are shown in Fig. S2.

**Spin-triplet MR measurement**

The MR measurement of the quasi-2D vdW $Fe_{0.29}TaS_2$/SC devices was performed in a Physical Properties Measurement System (PPMS; Quantum Design). A bias ($V_{bias}$) was applied between the SC electrode and one normal Pt electrode using a Keithley K2400, the source-drain current ($I_{sd}$) was measured using the same K2400, and the voltage ($V_{3T}$) between the SC electrode and the other Pt electrode was measured using a Keithley 2002. The interfacial resistance was obtained via dividing the three-terminal voltage by the source-drain current ($R_{3T} = V_{3T}/I_{sd}$). During the measurement of each spin-triplet MR curve, the quasi-2D vdW $Fe_{0.29}TaS_2$/SC device was rotated from 0 to 360 degrees under the external static magnetic field in the PPMS.

**Data availability**

The data that support the findings of this study are available from the corresponding authors upon request.

**References:**

1   Tokura, Y., Kawasaki, M. & Nagaosa, N. Emergent functions of quantum materials. *Nat. Phys.* **13**, 1056–1068, (2017).

2   Basov, D. N., Averitt, R. D. & Hsieh, D. Towards properties on demand in quantum materials. *Nat. Mater.* **16**, 1077–1088, (2017).

3   Bergeret, F. S., Volkov, A. F. & Efetov, K. B. Odd triplet superconductivity and related phenomena in superconductor-ferromagnet structures. *Rev. Mod. Phys.* **77**, 1321-1373, (2005).

4   Buzdin, A. I. Proximity effects in superconductor-ferromagnet heterostructures. *Rev. Mod. Phys.* **77**, 935-976, (2005).

5   Linder, J. & Robinson, J. W. A. Superconducting spintronics. *Nat. Phys.* **11**, 307-315, (2015).

6   Eschrig, M. Spin-polarized supercurrents for spintronics: a review of current progress. *Rep. Prog. Phys.* **78**, 104501, (2015).

7   Jeon, K.-R., Ciccarelli, C., Ferguson, A. J., Kurebayashi, H., Cohen, L. F., Montiel, X., Eschrig, M., Robinson, J. W. A. & Blamire, M. G. Enhanced spin pumping into superconductors provides evidence for superconducting pure spin currents. *Nat. Mater.* **17**, 499–503, (2018).

8   Han, W., Maekawa, S. & Xie, X.-C. Spin current as a probe of quantum materials. *Nat. Mater.* **19**, 139–152, (2020).

9   Sau, J. D., Lutchyn, R. M., Tewari, S. & Das Sarma, S. Generic New Platform for Topological Quantum Computation Using Semiconductor Heterostructures. *Phys. Rev. Lett.* **104**, 040502, (2010).

10  Aasen, D. et al. Milestones toward Majorana-based quantum computing, Phys. Rev. X 6, 031016 (2016).

11  Visani, C., Sefrioui, Z., Tornos, J., Leon, C., Briatico, J., Bibes, M., Barthélémy, A., Santamaría, J. & Villegas, J. E. Equal-spin Andreev reflection and long-range coherent transport in high-temperature superconductor/half-metallic ferromagnet junctions. *Nat. Phys.* **8**, 539-543, (2012).

12  Martínez, I., Högl, P., González-Ruano, C., Cascales, J. P., Tiusan, C., Lu, Y., Hehn, M., Matos-Abiague, A., Fabian, J., Žutić, I. & Aliev, F. G. Interfacial Spin-Orbit Coupling: A Platform for Superconducting Spintronics. *Phys. Rev. Appl.* **13**, 014030, (2020).

13  Bergeret, F. S. & Tokatly, I. V. Spin-orbit coupling as a source of long-range triplet proximity effect in superconductor-ferromagnet hybrid structures. *Phys. Rev. B* **89**, 134517, (2014).11

**Acknowledgements**

We thank E. I. Rashba for valuable discussions. R.C., Y.Y., Y.M., W.X., B.L., Y.J., H.Z., S.J., X.C.X., Q.-F.S., and W.H. acknowledge the financial support from National Basic Research Programs of China (Nos. 2019YFA0308401, 2017YFA0303301, and 2018YFA0305601), National Natural Science Foundation of China (Nos. 11974025, 11921005, 11947053, 11774007, and U1832214), Beijing Natural Science Foundation (No. 1192009), and the Key Research Program of the Chinese Academy of Sciences (Grant No. XDB28000000). P.L. also acknowledges the financial support from Hubei Provincial Natural Science Foundation of China (No. 2019CFB218). C.S. and I.Z. acknowledge the support from U.S. Department of Energy, Office of Science, Basic Energy Sciences under Award No. DE- SC0004890.

**Author contributions**

R.C., W.X., and B.L. prepared the 2D van der Waals ferromagnet $Fe_{0.29}TaS_2$ flakes and fabricated the devices. R.C., Y.Y., Y.M. and Y.J. performed the MR measurements. P.L., C.S., X.C.X., I.Z., and Q.-F.S. performed the theoretical analysis. H.Z. and S.J. synthesized the bulk $Fe_{0.29}TaS_2$ single crystals. W.H. supervised the experiments and wrote the manuscript with contributions from all authors. All the authors discussed the results and contributed to the final version of the manuscript.

**Competing Interests**

The authors declare no competing interests.

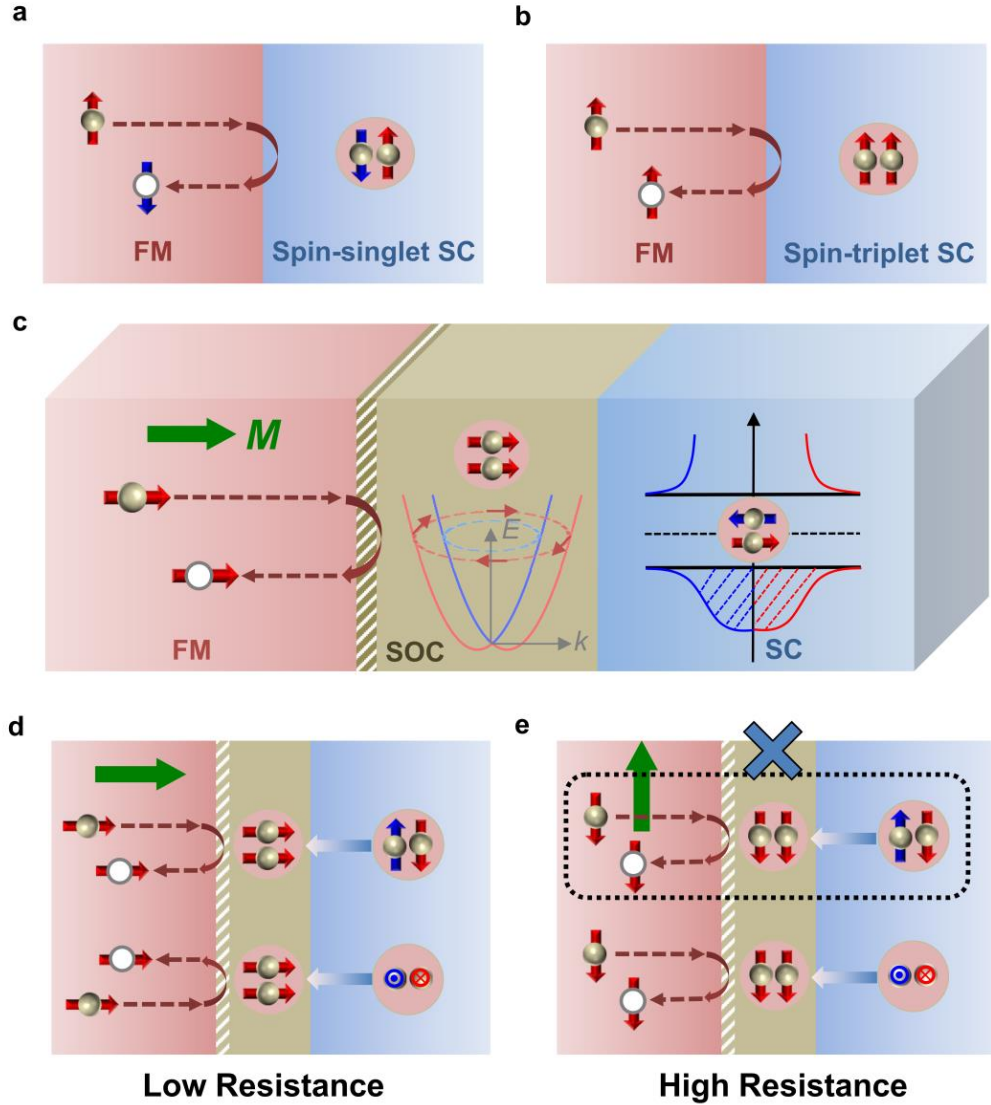

**Figure 1. Schematic of the spin-triplet Andreev reflection at FM/SC interface. a,** Conventional Andreev reflection at the FM/spin-singlet SC interface. **b,** The spin-triplet Andreev reflection at the FM/spin-triplet SC interface. **c,** Schematic of the spin-triplet Andreev reflection resulting from Rashba SOC at the interface between a FM and a conventional *s*-wave SC. The arrows in Rashba SOC band indicate spin-momentum locking and the red arrows represent the spin-polarization direction of equal-spin-triplet pairs. **d-e,** Anisotropic spin-triplet Andreev reflection at the FM/SC interface and the low/high interfacial resistance states that depend on the FM magnetization direction, **M** (green arrow). Red arrows at the interface denote the spin direction of equal-spin-triplet pairs. For **M** along the interface the spin-triplet Andreev reflection can be suppressed.

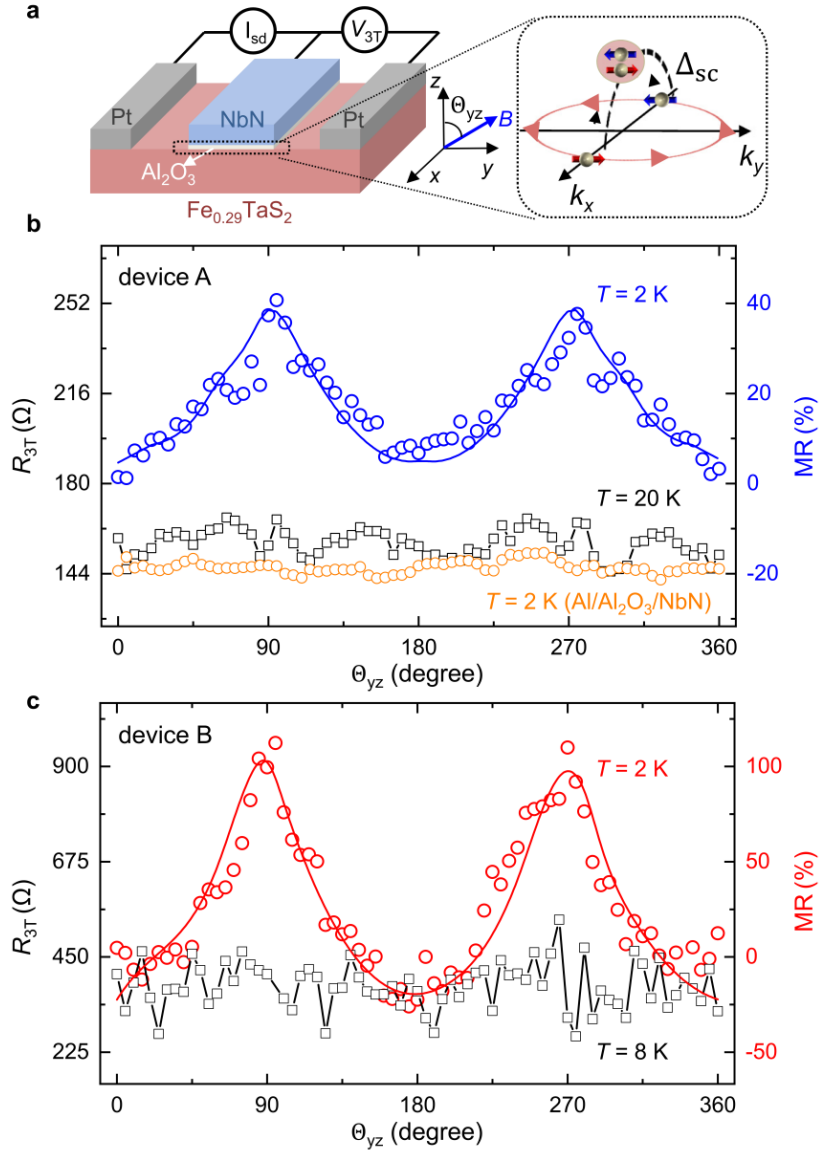

**Figure 2. Large magnetoresistance of the quasi-2D vdW Fe$_{0.29}$TaS$_2$/SC junction. a,** Illustration of the quasi-2D vdW Fe$_{0.29}$TaS$_2$/SC MR device and the measurement geometry. The right panel shows the schematic of the spin-triplet pairing component resulting from Rashba SOC at the FM/SC interface. **b,** The interfacial resistance ($R_{3T} = V_{3T}/I_{sd}$) and MR ratio as a function of the magnetic field angle measured on the typical quasi-2D vdW Fe$_{0.29}$TaS$_2$/SC device (device A) under $B = 9$ T. The orange curve represents the resistance measured on a typical control device (Al/Al$_2$O$_3$/NbN) under $B = 9$ T. **c,** The interfacial resistance and MR ratio as a function of the magnetic field angle on device B under $B = 9$ T. The solid lines in **b** and **c** are guides to the eye.

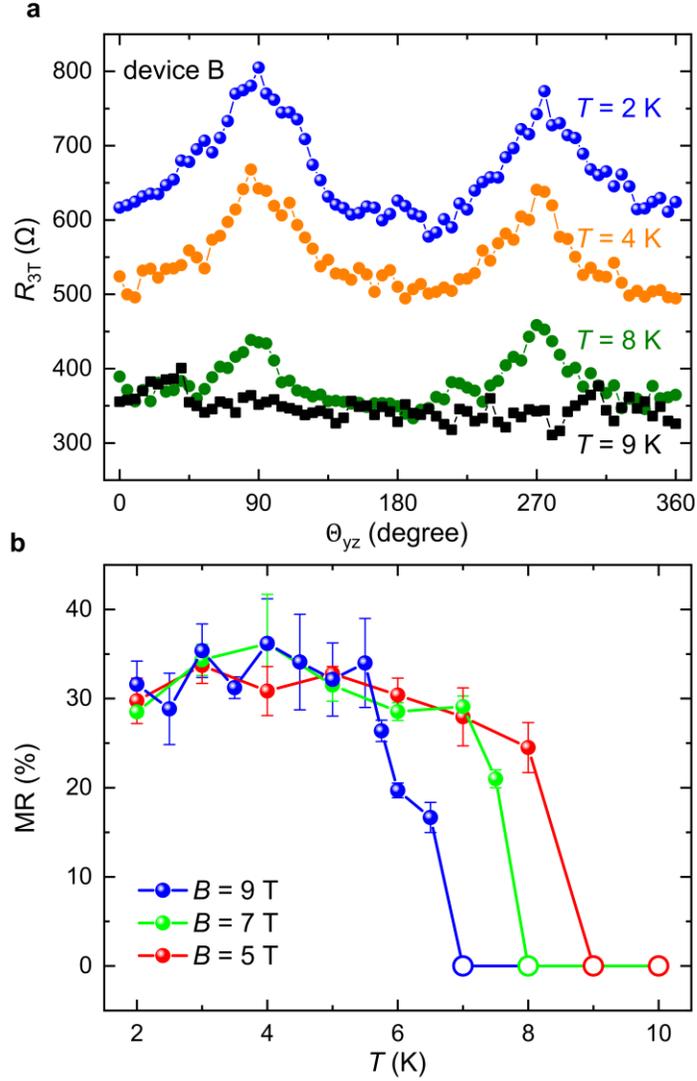

**Figure 3. The temperature dependence of MR at $Fe_{0.29}TaS_2$/SC interface. a,** The interfacial resistance as a function of $\Theta_{yz}$ measured on device B at $T = 2$ K (blue), 4 K (yellow) 8 K (olive), and 9 K (black), respectively. These results were obtained under $B = 5$ T and $V_{bias} = 1$ mV, which correspond to $V_{3T} \sim 0.40$ mV for $T = 2$ and 4 K, and $V_{3T} \sim 0.25$ mV for $T = 8$ and 9 K. **b,** The temperature dependence of MR ratio of device B at $B = 9$ T, 7 T, and 5 T, respectively. The error bars correspond to one standard deviation. The open circles represent the absence of obvious MR.

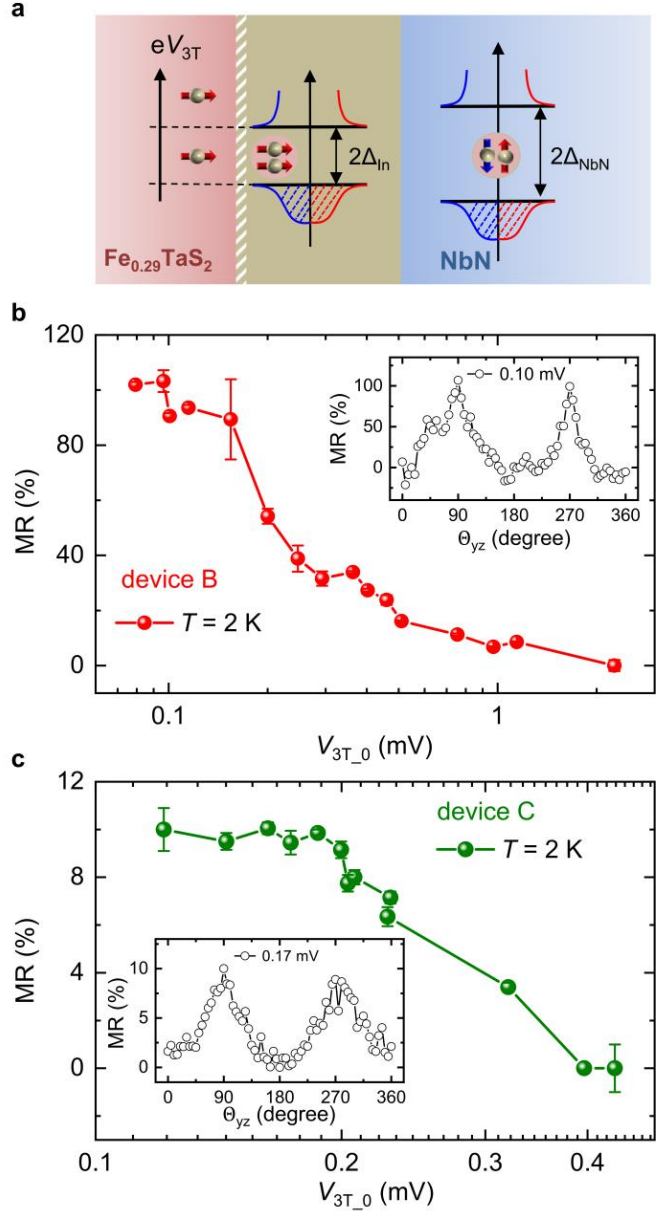

**Figure 4. The voltage dependence of MR at $Fe_{0.29}TaS_2$/SC interface. a,** Schematic of the incident spin-polarized electrons with chemical potentials inside and above the interface spin-triplet superconducting energy gap. $\Delta_{In}$ and $\Delta_{NbN}$ indicate the superconducting energy gaps of the interface SC and the bulk NbN. **b,** The voltage dependence ($V_{3T\_0}$) of the MR ratio of device B measured at $T = 2$ K and $B = 9$ T. $V_{3T\_0}$ represents $V_{3T}$ when an applied magnetic field is perpendicular to the FM/SC interface. The error bars correspond to one standard deviation. Inset: The typical MR curve at $V_{3T\_0} = 0.10$ mV. **c,** The voltage dependence ($V_{3T\_0}$) of the MR ratio of device C measured at $T = 2$ K and $B = 9$ T. The error bars correspond to one standard deviation. Inset: The typical MR curve at $V_{3T\_0} = 0.17$ mV.

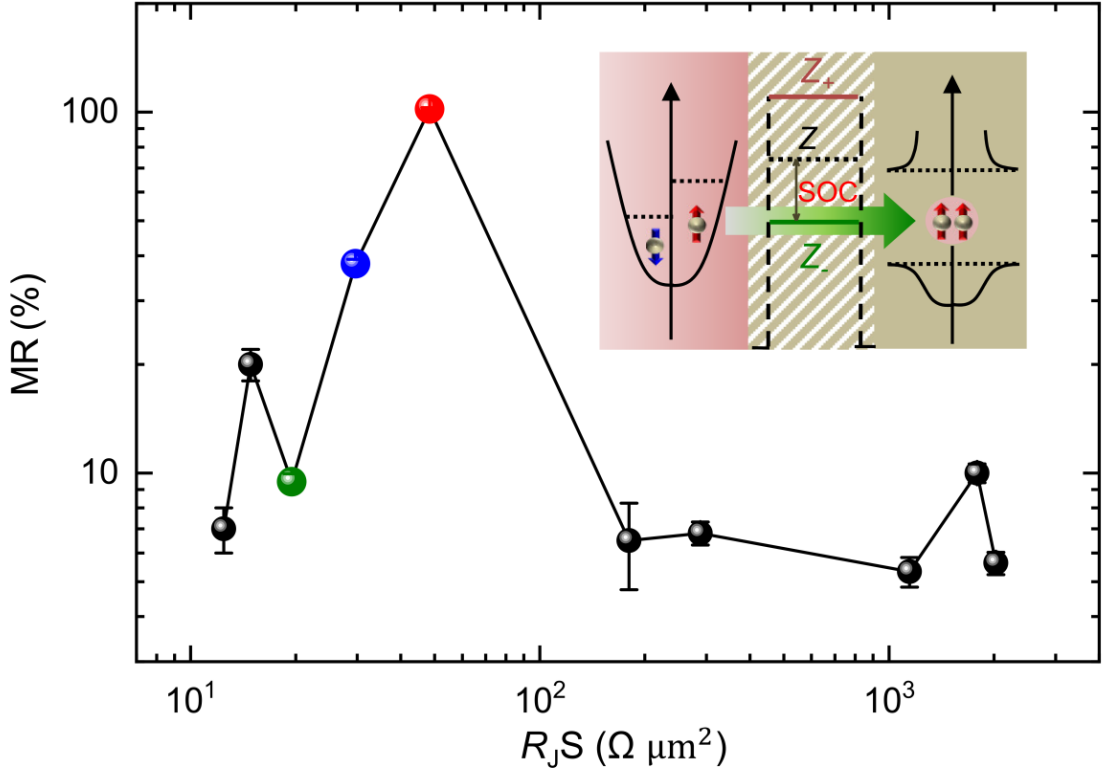

**Figure 5. The interface barrier dependence of MR at Fe$_{0.29}$TaS$_2$/SC interface.** The MR ratio as a function of the interface resistance area product ($R_J S$) measured on various devices in the low voltage bias region. Inset: Schematic of the incident spin-polarized electrons into the interfacial spin-triplet SC via interface barrier with Rashba SOC. The Rashba SOC modifies the interface barrier strength ($Z$) to be $Z_\pm = Z \pm \bar{\gamma} k_\parallel$, where $\bar{\gamma}$ is the SOC parameter and $k_\parallel$ is the in-plane wave vector[25]. The blue, red, and green dots represent the MR of devices A, B, and C, respectively. The error bars correspond to one standard deviation.

# Supplementary Information for

# Evidence for spin-triplet Andreev reflection at the 2D van der Waals ferromagnet/superconductor interface

Ranran Cai[1,2,†], Yunyan Yao[1,2,†], Peng Lv[3], Yang Ma[1,2], Wenyu Xing[1,2], Boning Li[1,2], Yuan Ji[1,2], Huibin Zhou[1,2], Chenghao Shen[4], Shuang Jia[1,2,5,6], X. C. Xie[1,2,5,6], Igor Zutic[4], Qing-Feng Sun[1,2,5,6] & Wei Han[1,2]*

[1]International Center for Quantum Materials, School of Physics, Peking University, Beijing 100871, P. R. China
[2]Collaborative Innovation Center of Quantum Matter, Beijing 100871, P. R. China
[3]Department of Physics, Wuhan University of Technology, Wuhan 430070, China
[4]Department of Physics, University at Buffalo, State University of New York, Buffalo, New York 14260, USA
[5]CAS Center for Excellence in Topological Quantum Computation, University of Chinese Academy of Sciences, Beijing 100190, P. R. China
[6]Beijing Academy of Quantum Information Sciences, Beijing 100193, P. R. China
[†]These authors contributed equally to the work
*Correspondence to: weihan@pku.edu.cn (W.H.)

**This file includes the following:**

1) Supplementary Note S1-S4
2) Supplementary References
3) Table S1
4) Figures S1-S10

**Supplementary Note S1. Spin-triplet pairings induced by interfacial spin-orbit couplings**

Due to the lack of inversion symmetry at the interface between a ferromagnet (FM) and an *s*-wave superconductor (SC), there are generally two types of spin-orbit coupling (SOC), namely Rashba and Dresselhaus SOC. Close to the interface, the spin-momentum locking property of SOC breaks the spin-rotation symmetry, leading to spin-triplet Cooper pairs and spin-triplet Andreev reflection at the FM/SC interface. In this section, we calculate the correlation functions of the superconductor with both Rashba and Dresselhaus SOC. Consider the normal-state Hamiltonian of the superconductor as follows[1],

$$h(\mathbf{k}) = \sum_{\mathbf{k},s} \varepsilon_{\mathbf{k}} c^{\dagger}_{\mathbf{k}s} c_{\mathbf{k}s} + \sum_{\mathbf{k},ss'} [\alpha_R(k_y\sigma_x - k_x\sigma_y) + \alpha_D(k_x\sigma_x - k_y\sigma_y)]_{ss'} c^{\dagger}_{\mathbf{k}s} c_{\mathbf{k}s'}, \quad (1)$$

where $\varepsilon_{\mathbf{k}} = \frac{\hbar^2 \mathbf{k}^2}{2m_S} - \mu_S$ is the single-electron kinetic energy measured from the chemical potential $\mu_S$, $m_S$ is the electron mass in SC and $\hbar$ is the reduced Plank constant, $c^{\dagger}_{\mathbf{k}s}(c_{\mathbf{k}s'})$ is the creation (annihilation) operator of electron, $s, s' = \uparrow, \downarrow$ are the spin indices, $\alpha_R$ and $\alpha_D$ are the Rashba and Dresselhaus SOC strength parameters, respectively, $\mathbf{k} = (k_x, k_y, k_z)$ is the three-dimensional electron wave vector, and $(\sigma_x, \sigma_y)$ are the Pauli matrices in the *x-y* plane. Assuming the usual spin-singlet *s*-wave paring, the mean-field Bogoliubov-de Gennes (BdG) Hamiltonian of Eq. (S1) in the Nambu basis $(c_{\mathbf{k}\uparrow}, c_{\mathbf{k}\downarrow}, c^{\dagger}_{-\mathbf{k}\uparrow}, c^{\dagger}_{-\mathbf{k}\downarrow})$ can be written as

$$H(\mathbf{k}) = \begin{pmatrix} h(\mathbf{k}) & i\sigma_y \Delta \\ -i\sigma_y \Delta & -h^*(-\mathbf{k}) \end{pmatrix}. \quad (2)$$

The in-plane SOC breaks the spin-rotation symmetry of $h(\mathbf{k})$, and then the *s*-wave pairing matrix $i\sigma_y \Delta$ in Eq. (S2) can induce spin-triplet Cooper pairs. This can be shown by studying the pairing symmetry. After solving the standard Gor'kov equations[2,3], the corresponding pairing correlations can be written as

$$\mathbf{F}(\mathbf{k}, E) = \Delta[d_0(\mathbf{k}, E)\sigma_0 + \mathbf{d}(\mathbf{k}, E) \cdot \boldsymbol{\sigma}]i\sigma_y. \quad (3)$$

where $d_0(\mathbf{k}, E) = F_+(\mathbf{k}, E)$ characterizes the spin-singlet part, $\mathbf{d}(\mathbf{k}, E) = F_-(\mathbf{k}, E)\hat{\mathbf{g}}_\mathbf{k}$ characterizes the spin-triplet part, $E$ is the energy eigenvalue, $\mathbf{g}_\mathbf{k} = (\alpha_R k_y + \alpha_D k_x, -(\alpha_R k_x + \alpha_D k_y), 0)$ is the SOC field with unit vector $\hat{\mathbf{g}}_\mathbf{k} = \mathbf{g}_\mathbf{k}/|\mathbf{g}_\mathbf{k}|$, and

$$F_\pm(\mathbf{k}, E) = \frac{1}{2}\left[\frac{1}{E^2 - (\varepsilon_\mathbf{k} + |\mathbf{g}_\mathbf{k}|)^2 - \Delta^2} \pm \frac{1}{E^2 - (\varepsilon_\mathbf{k} - |\mathbf{g}_\mathbf{k}|)^2 - \Delta^2}\right]. \quad (4)$$

In the absence of the SOC field (i.e., $|\mathbf{g}_\mathbf{k}| = 0$), the spin-triplet part $\mathbf{d}(\mathbf{k}, E) = F_-(\mathbf{k}, E)\hat{\mathbf{g}}_\mathbf{k}$ vanishes, and the system exhibits conventional *s*-wave superconductivity. On the other hand, in the presence of the SOC with the nonzero $\mathbf{g}_\mathbf{k}$, the system supports mixed *s*-wave and *p*-wave superconductivity, with the spin-triplet part being linearly proportional to $|\mathbf{g}_\mathbf{k}|$ for small SOC field. It is important to note that in the basis where the spin-quantization axis is along the out-of-plane direction (*z*-direction, see Fig. 2a in the main text), only $d_x, d_y$ components of the spin-triplet part $\mathbf{d}(\mathbf{k}, E)$ are nonzero, and all the spin-triplet Cooper pairs are formed by electrons with equal spins (Table. S1). Hence, besides conventional Andreev reflection, spin-triplet Andreev reflection also occur at the FM/SC interface, as shown in Fig. S1a. Here, we emphasize that $\sqrt{d_x^2 + d_y^2}$ is nonzero for arbitrary in-plane wave vector $(k_x, k_y)$, which indicates that the spin-triplet Andreev reflection can occur without any constraint if the spins of the incident electrons are along the out-of-plane direction.

For incident electrons with spins pointing along the *xy* plane, the spin-triplet Andreev reflection can be suppressed for certain in-plane wave vectors. For example, considering the spins of incident electrons being along the $x$ axis, $\sqrt{d_y^2 + d_z^2} = |\alpha_R k_x + \alpha_D k_y|$ can be zero at some special wave vectors $(k_x, k_y)$, leading to the disappearance of the spin-triplet Andreev reflection. To see this more clearly, we introduce an angle $\vartheta$ to denote the strength ratio between Dresselhaus and Rashba SOC, with $\alpha_R = \gamma\cos\vartheta$, $\alpha_D = \gamma\sin\vartheta$, $\gamma = \sqrt{\alpha_R^2 + \alpha_D^2}$. We also introduce an angle $\varphi$ to denote the azimuth of in-plane wave vector $k_\parallel$, with $k_x = k_\parallel\cos\varphi$, $k_y = k_\parallel\sin\varphi$, $k_\parallel = \sqrt{k_x^2 + k_y^2}$. Then the SOC field $\mathbf{g_k}$ can be simplified to $(k_\parallel\gamma\sin(\varphi + \vartheta), -k_\parallel\gamma\cos(\varphi - \vartheta), 0)$. By defining $\sin\rho = \sin(\varphi + \vartheta)/A$, $\cos\rho = \cos(\varphi - \vartheta)/A$, $A = \sqrt{1 + \sin2\vartheta\sin2\varphi}$, the pairing correlations in Eq. (S3) can be rewritten as

$$\mathbf{F}(\mathbf{k}, E) = \Delta \begin{pmatrix} -ie^{-i\rho} F_- & F_+ \\ -F_+ & -ie^{i\rho} F_- \end{pmatrix}. \tag{5}$$

By choosing a new spin-quantization axis in the *x-y* plane, the pairing correlations become

$$\tilde{\mathbf{F}}(\mathbf{k}, E) = \frac{\Delta}{2}\begin{pmatrix} -ie^{-i\rho}\left[e^{2i(\rho-\phi)} + 1\right]F_- & 2F_+ + ie^{i(\phi-\rho)}\left[1 - e^{2i(\rho-\phi)}\right]F_- \\ -2F_+ + ie^{i(\phi-\rho)}\left[1 - e^{2i(\rho-\phi)}\right]F_- & -ie^{i\rho}\left[e^{2i(\phi-\rho)} + 1\right]F_- \end{pmatrix}, \tag{6}$$

where $\phi$ is the angle between the spin-quantization axis and the $x$ axis. The spin-triplet Andreev reflection arises from the diagonal part of the paring correlations, which vanishes if

$$e^{2i(\rho-\phi)} = -1 \Rightarrow \rho = \phi \pm \pi/2, \tag{7}$$

and reaches a maximum if

$$e^{2i(\rho-\phi)} = 1 \Rightarrow \rho = \phi, \phi + \pi. \tag{8}$$

Note that the parameter $\rho$ encodes the in-plane wave vector $(k_x, k_y)$ of the spin-polarized electrons, which means that for certain in-plane wave vectors, the spin-triplet Andreev reflection is strictly prohibited. Compared with the case where the electrons' spin directions are out-of-plane, we can infer that the conductance is weaker when the spin directions of incident electrons are in-plane. As an example, consider the spin direction of incident electrons and the spin-quantization along the $x$ axis (i.e., $\phi = 0$).

In the absence of the Dresselhaus SOC ($\vartheta = 0$), we have $\rho = \varphi$ from the definition of $\rho$. The minimum condition in Eq. (S7) now reduces to

$$\varphi = \pm\pi/2 \Rightarrow k_x = 0, k_y \neq 0 , \qquad (9)$$

which means for incident electrons with in-plane wave vector $(0, k_y)$, the spin-triplet Andreev reflection is strictly prohibited, since the Cooper pairs are completely formed by electrons with opposite spins, as illustrated in Fig. S1b. On the other hand, the maximum condition in Eq. (S8) now reduce to

$$\varphi = 0, \pi \Rightarrow k_x \neq 0, k_y = 0 , \qquad (10)$$

which means for incident electrons with in-plane wave vector $(k_x, 0)$, the spin-triplet Andreev reflection is the most significant as shown in Fig. S1c.

**Supplementary Note S2. Characterization of quasi-2D vdW FM $Fe_{0.29}TaS_2$ via anomalous Hall effect**

The magnetic properties of quasi-2D vdW FM $Fe_{0.29}TaS_2$ are measured via anomalous Hall effect (AHE) on the Hall bar devices (Figs. S4 and S5). The Curie temperature is obtained to be ~ 90 K at the temperature when the anomalous Hall resistance disappears (Fig. S4d). As shown in Fig. S4c and Fig. S5a, the magnetization (*M*) easy axis is perpendicular to the $Fe_{0.29}TaS_2$ plane. The angle between *M* and the interface normal ($\theta_M$) can be estimated using the evolution of AHE with applied magnetic field, from $\theta_M = \arccos(R_{AHE}^{B_{IP}}/R_{AHE}^{B_\perp})$, where $B_{IP}$ and $B_\perp$ represent the in-plane and out-of-plane magnetic fields and they label the corresponding AHE resistance. The magnetic field dependence of $\theta_M$ is shown in Fig. 5b.

Figure S7 shows the MR results as a function of the extenal magnetic field measured on device B. Clearly, similar trends of the MR and $\theta_M$ are observed as a function of the magnetic field. This observation suggests that the measured MR is *M*-dependent, which is consistent with the strong anisoptric spin-triplet Andreev refelction at the $Fe_{0.29}TaS_2$/SC interface.

**Supplementary Note S3. Bias dependence of the conductance at the $Fe_{0.29}TaS_2$/SC interface.**

The interface conductance is characterized by the bias dependence at various temperature and magnetic fields (Figs. S6). Around zero bias, the conductance at $T = 2$ K is about 0.8 of the normal state value, $G_N$, which indicates that the interface barrier has a modest strength, rather than a strong barrier with a low transparency[4]. For such a modest interfacial strength as in our $Fe_{0.29}TaS_2$/SC junctions, the contribution of the Andreev reflection to the zero-bias interface conductance remains significant.

**Supplementary Note S4. Control experiments on the vortex-induced anisotropic transport in superconductor heterostructures**

In this section, we discuss the role of vortex-induced anisotropic interfacial resistance in the type-II SC heterostructures[5,6]. As it is well known that the formation of vortices in SC is strongly dependent on the direction and amplitude of the external mangetic field[6,7], the vortices in type-II SC might also contribute to the anisotropic transport properties in the FM/SC heterostructures. To study the role of the vortices and the anisotropic interfacial resistance purely from vortex in the SC NbN electrode, we fabricate control devices that use ~20 nm Al as a NM to replace the quasi-2D vdW FM Fe$_{0.29}$TaS$_2$. As seen in Fig. S8, The only difference between the spin-triplet MR device (Fig. S8a) and control device (Fig. S8b) is the bottom layer; NM Al *vs.* quasi-2D vdW FM Fe$_{0.29}$TaS$_2$. These two devices are chosen for the comparision due to similar values of the interface resistance area product ($R_JS$) of 29.7 and 45.6 Ω μm$^2$ for spin-triplet MR device and control device, respectively.

Using the same measurement geomery and under the same conditions ($T$ = 2 K, $B$ = 9 T, and $V_{bias}$ = 1 mV) as the spin-triplet MR device, the angle dependence of the interfacial resistance between the Al electrode and the NbN electrode is measured. Clearly, the vortex-induced MR in control device (red symbols in Fig. S8c) is significantly smaller compared to the spin-triplet MR (blue symbols in Fig. S8c). Furthermore, the MR in the control device is within the noise level of ~3% (Fig. S8d). To rule out any significant contribution from vortex in SC, the control devices with various $R_JS$ have been fabricated and measured. The signals of these control devices are summarized in Fig. S8e (red symbols), which are significantly smaller compared to the MR at quasi-2D vdW Fe$_{0.29}$TaS$_2$/SC interface (blue symbols). To conclude, the contribution of the vortices in type-II SC to the interface transport properties is negligible compared to the spin-triplet MR in our study on the spin-triplet Andreev reflection of the quasi-2D vdW FM Fe$_{0.29}$TaS$_2$/SC devices.

**Supplementary References:**

| $d_{0,x,y,z}$ | pairing symmetry | value | type of AR |
|---|---|---|---|
| $d_0$ | singlet | $1/2\langle c_{k\uparrow}c_{-k\downarrow} - c_{k\downarrow}c_{-k\uparrow}\rangle$ | conventional AR |
| $d_x$ | triplet | $-1/2\langle c_{k\uparrow}c_{-k\uparrow} - c_{k\downarrow}c_{-k\downarrow}\rangle$ | spin-triplet AR |
| $d_y$ | triplet | $-i/2\langle c_{k\uparrow}c_{-k\uparrow} + c_{k\downarrow}c_{-k\downarrow}\rangle$ | spin-triplet AR |
| $d_z$ | triplet | $1/2\langle c_{k\uparrow}c_{-k\downarrow} + c_{k\downarrow}c_{-k\uparrow}\rangle$ | conventional AR |

**Table S1: A summary list of the pairing functions of mixed *s*- and *p*-wave superconductivity.** The spins of incident electrons are chosen to be along the *z* axis. The singlet part $d_0$ and the triplet part $d_z$ are formed by electrons with opposite spins along the *z* direction, while $d_x$ and $d_y$ are formed by electrons with equal spins. The nonzero $d_x$ or $d_y$ gives rise to unconventional spin-triplet Andreev reflection (AR).

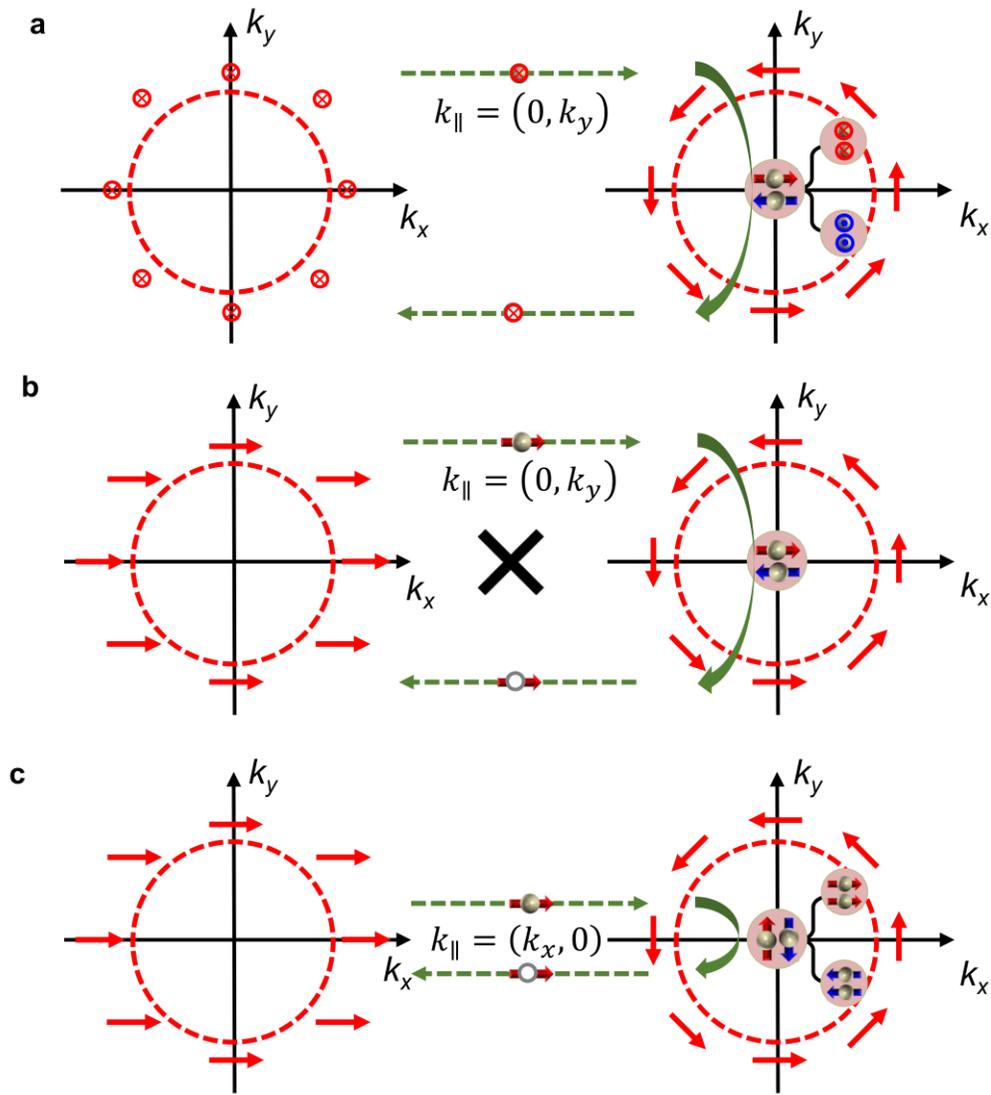

**Figure S1. The schematic diagram of the spin-triplet Andreev reflection at the FM/SC interface. a,** For incident electrons with out-of-plane spins, the spin-triplet Andreev reflection can occur for arbitrary in-plane wave vector ($k_x$, $k_y$), due to the formation of equal-spin Cooper pairs. **b, c,** For incident electrons with spins along the *x* axis, the spin-triplet Andreev reflection cannot occur for in-plane wave vector (0, $k_y$), but is significant for in-plane wave vector ($k_x$, 0).

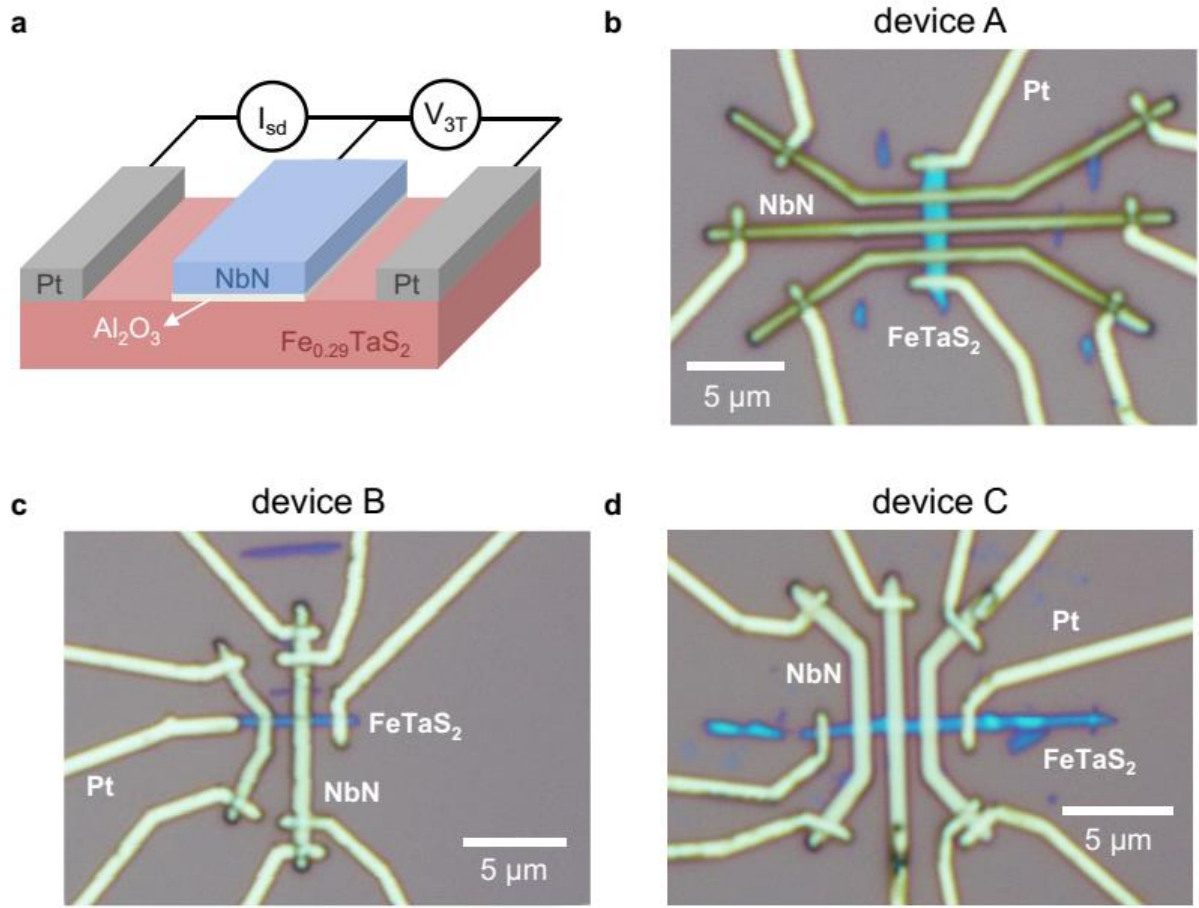

**Figure S2. The optical images of the representative devices. a,** Illustration of the quasi-2D vdW $Fe_{0.29}TaS_2$/SC MR device and the measurement geometry. Between the SC and the $Fe_{0.29}TaS_2$ flake, a thin $Al_2O_3$ layer (~ 1 - 2.5 nm) is used to tune the interface coupling strength. **b,** The optical image of Device A. Three SC NbN electrodes are fabricated onto the central part of the $Fe_{0.29}TaS_2$ flake, and two normal metal Pt electrodes are contacted on the two ends of the $Fe_{0.29}TaS_2$ flake. The thickness of the $Fe_{0.29}TaS_2$ flake is estimated to be ~ 20 nm. **c,** The optical image of device B. The thickness of the $Fe_{0.29}TaS_2$ flake is estimated to be ~ 15 nm. **d,** The optical image of device C. The thickness of the $Fe_{0.29}TaS_2$ flake is estimated to be ~ 20 nm.

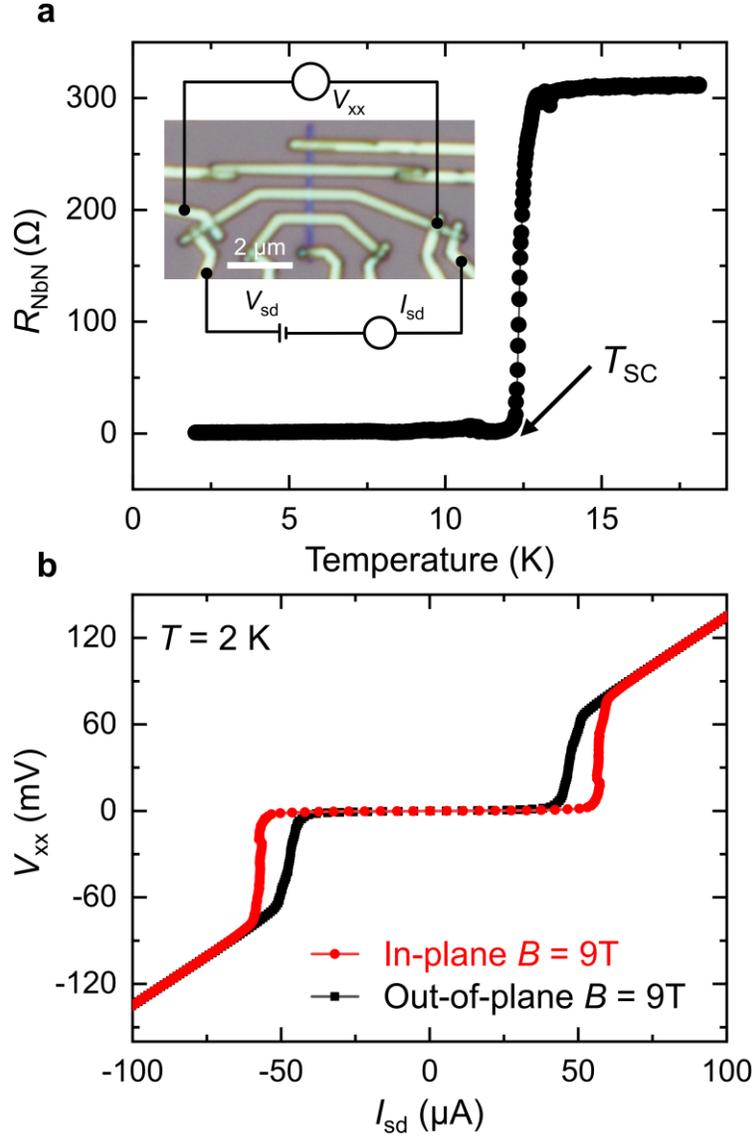

**Figure S3. Characterization of the SC electrodes**. **a**, The resistance of the NbN electrode as a function of the temperature. The superconducting critical temperature ($T_{SC}$) is determined to be ~ 12.5 K. Inset: The characterization of the SC electrode's resistance on a typical device using the standard four-probe measurement geometry. **b**, The current-voltage characteristics of the NbN electrode measured at $T = 2$ K under the in-plane (red) and out-of-plane (black) magnetic fields ($B = 9$ T), respectively.

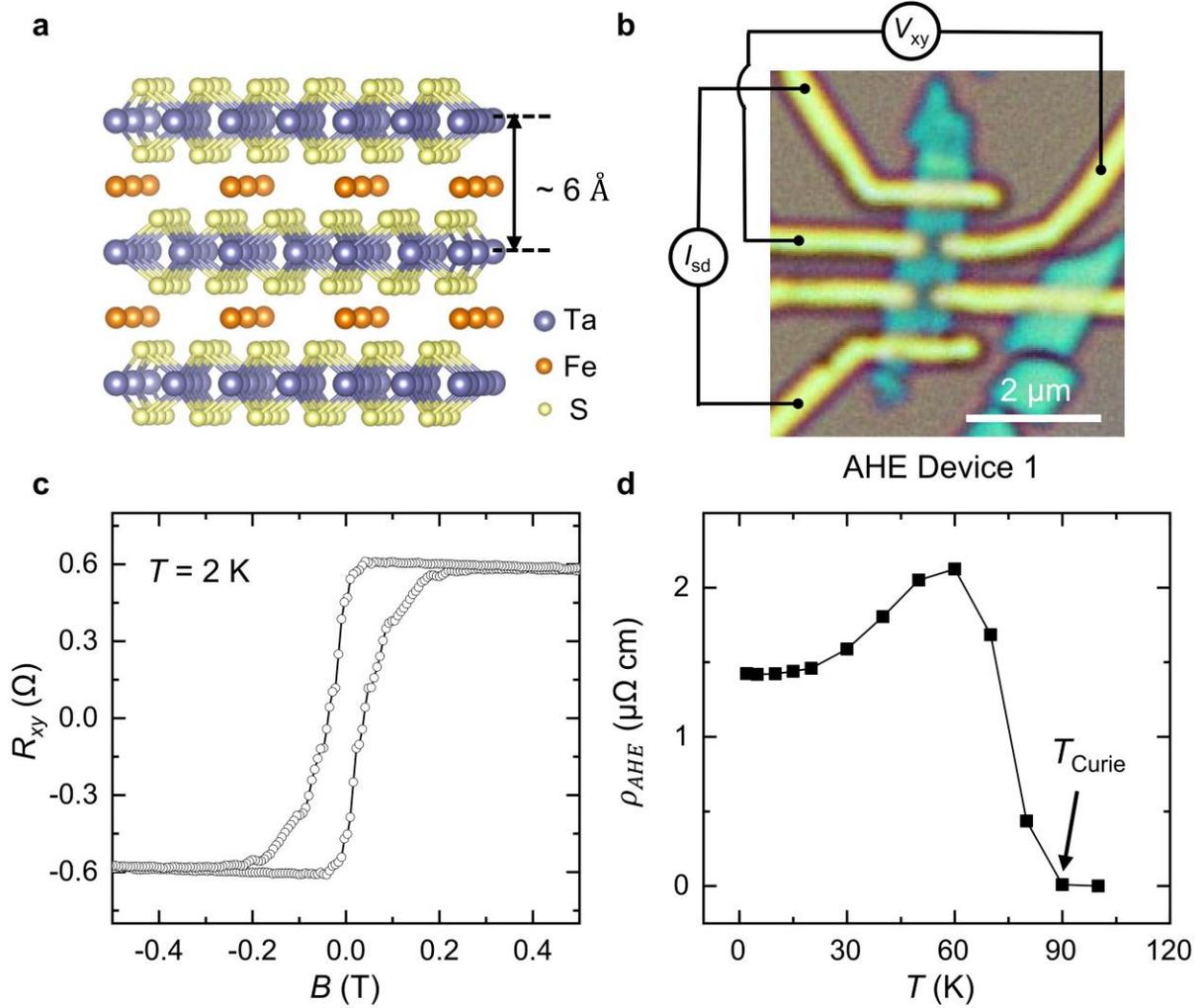

**Figure 4. Magnetic properties of quasi-2D vdW Fe$_{0.29}$TaS$_2$. a,** The side view of the crystal structure of itinerant quasi-2D vdW FM Fe$_{0.29}$TaS$_2$ flakes. The Fe atoms are located between the TaS$_2$ layers, which are stacked together *via* vdW interaction with an interlayer distance of ~ 6 Å. **b,** The optical image of the Fe$_{0.29}$TaS$_2$ anomalous Hall effect (AHE) device and the measurement geometry. The thickness of the quasi-2D vdW Fe$_{0.29}$TaS$_2$ is ~ 14 nm determined *via* atomic force microscopy. **c,** The transverse resistance ($R_{xy}$) as a function of the out-of-plane magnetic field measured on the Fe$_{0.29}$TaS$_2$ AHE device at $T$ = 2 K. **d,** The temperature-dependent anomalous Hall resistivity of the quasi-2D vdW Fe$_{0.29}$TaS$_2$ flake. The Curie temperature ($T_{Curie}$) is ~ 90 K, indicated by the black arrow.

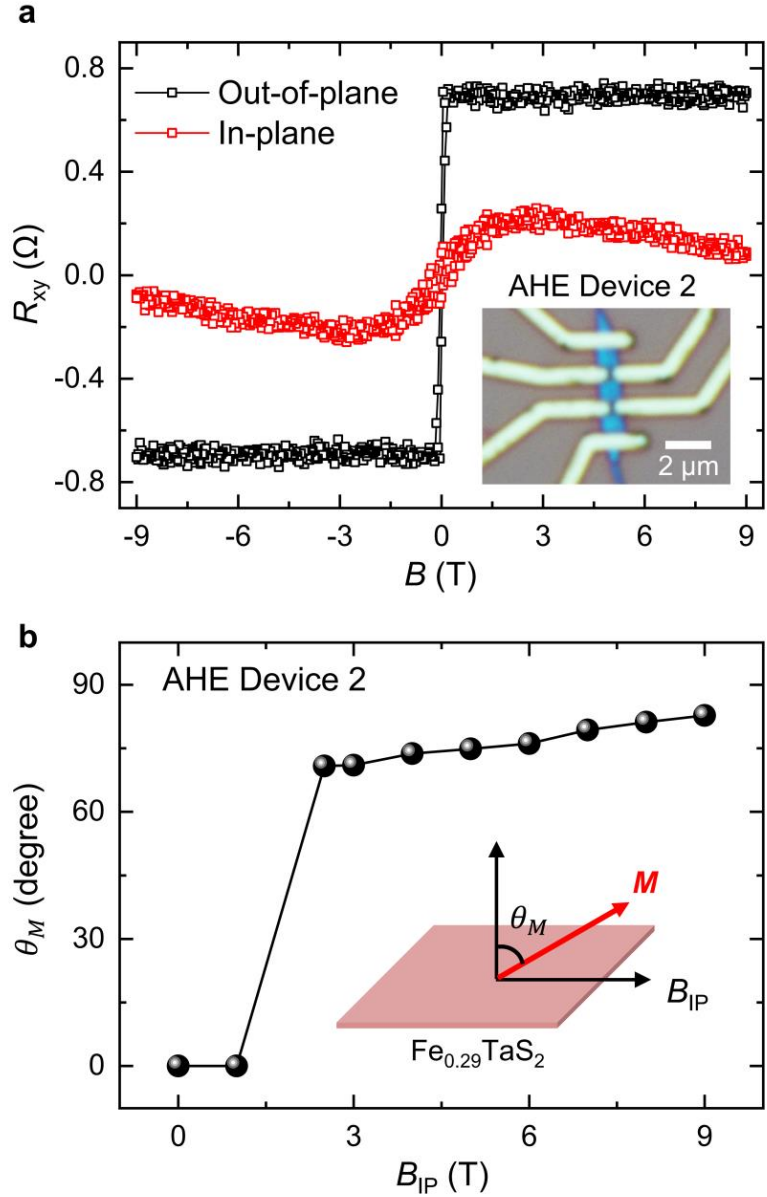

**Figure S5. Magnetization angle characterization of quasi-2D vdW $Fe_{0.29}TaS_2$ via anomalous Hall effect. a,** The transverse resistance ($R_{xy}$) as a function of the out-of-plane and in-plane magnetic fields measured on the $Fe_{0.29}TaS_2$ AHE device at $T = 2$ K. Inset: The optical image of the $Fe_{0.29}TaS_2$ AHE device. The thickness of the 2D vdW $Fe_{0.29}TaS_2$ is ~ 20 nm. **b,** The magnetization angle as a function of the in-plane magnetic field ($B_{IP}$). Inset: The schematic of $\theta_M$ under applied $B_{IP}$.

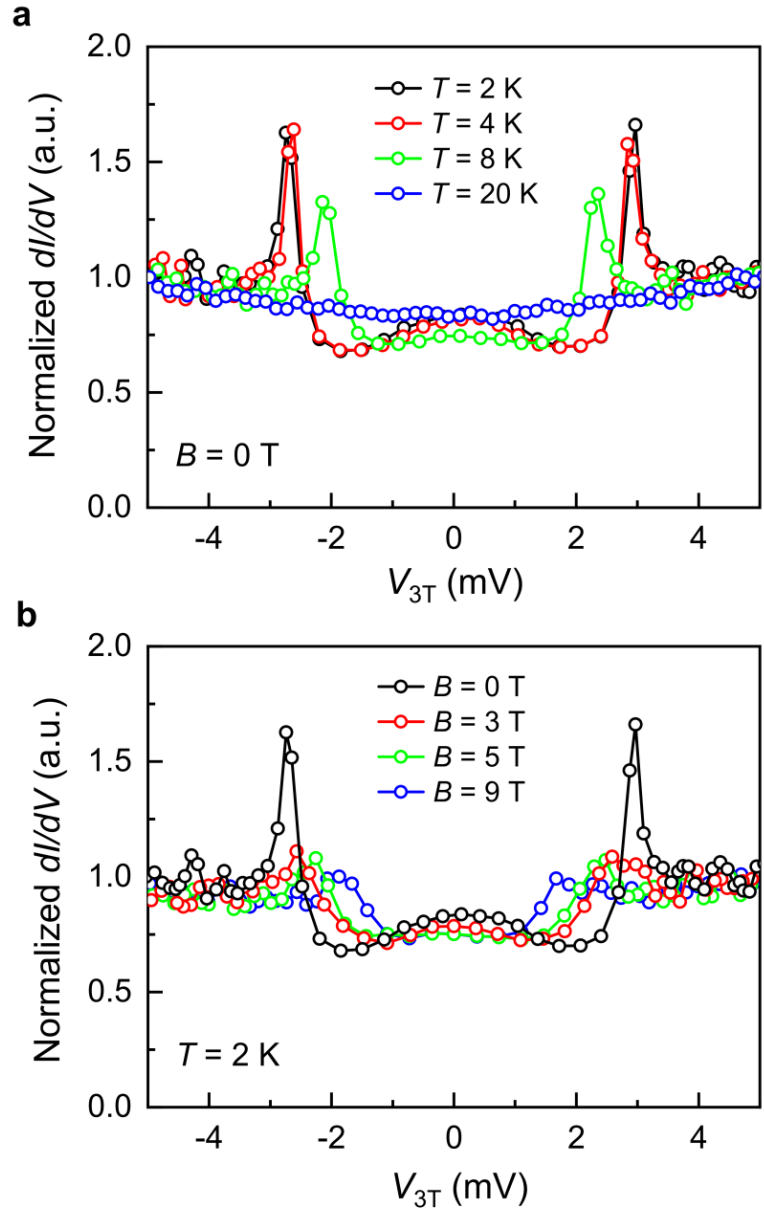

**Figure S6. Bias dependence of the conductance ($dI/dV$) at the Fe$_{0.29}$TaS$_2$/SC interface**. **a**, The $dI/dV$ curves at temperatures from $T = 2$ K to $T = 20$ K at $B = 0$ T. **b,** The $dI/dV$ curves at magnetic fields from $B = 0$ T to $B = 9$ T at $T = 2$ K. The results were obtained on the device C.

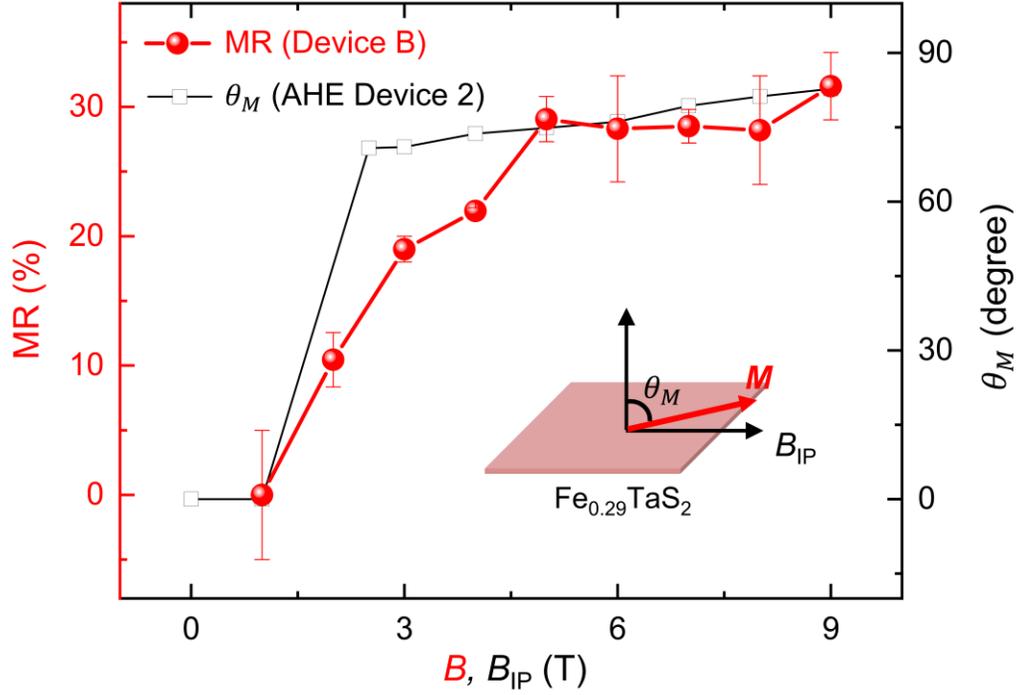

**Figure S7. The MR of Fe$_{0.29}$TaS$_2$/NbN as a function of the external magnetic field.** The red dots represent the MR measured on the device B at $T = 2$ K and $V_{3T} \sim 0.4$ mV, and the open black squares represent the magnetization angle as a function of the in-plane magnetic field at $T = 2$ K.

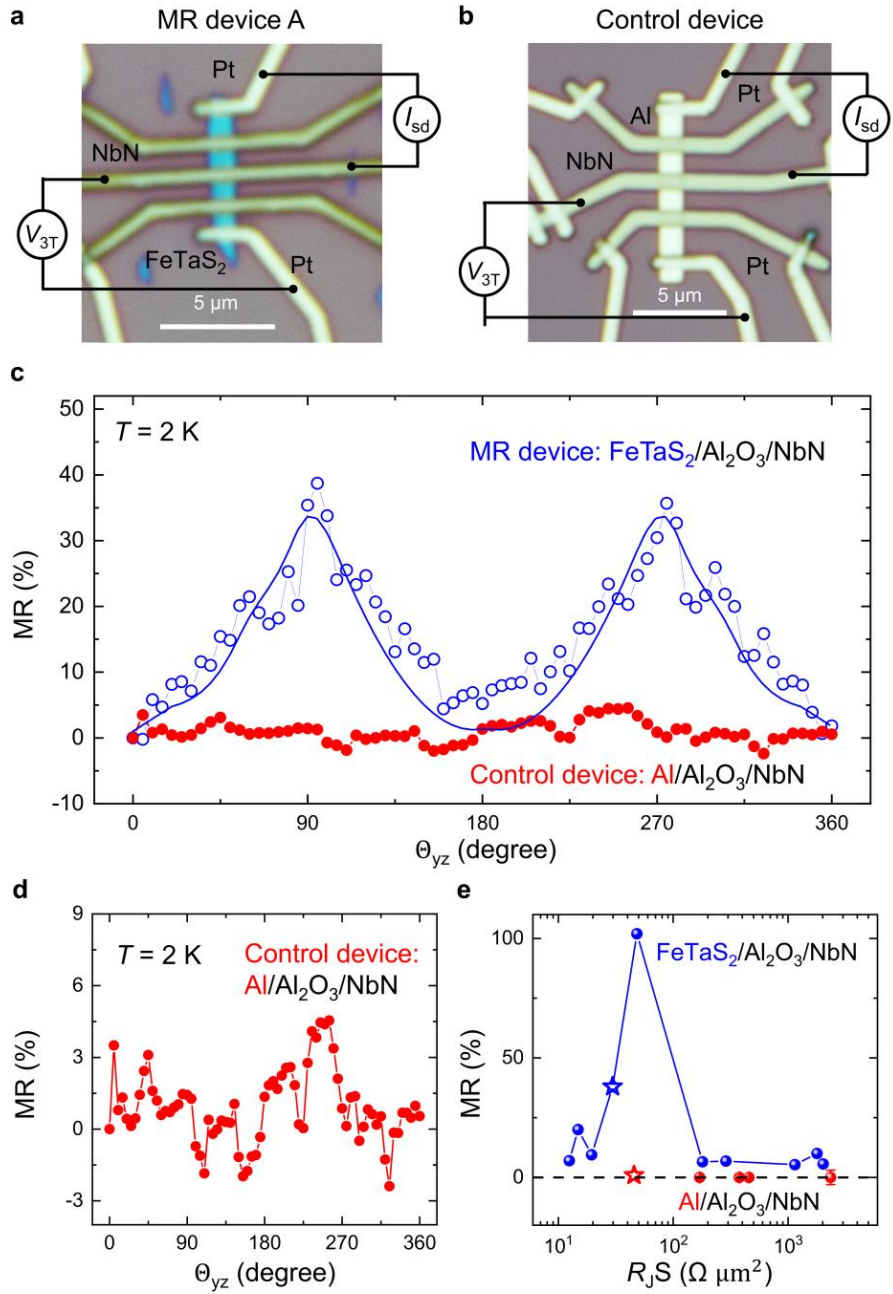

**Figure S8. Comparison of spin-triplet MR and SC vortex-induced MR. a, b,** The optical images of the spin-triplet MR device (device A) and the control device that measures the vortex-induced MR and their measurement geometry. For the control device, everything is the same as the spin-triplet MR device except that the 2D vdW FM $Fe_{0.29}TaS_2$ flake is replaced by a 20 nm Al electrode. **c,** The comparison of the spin-triplet MR (blue symbols) and vortex-induced MR (red

symbols) as a function of the magnetic field angle measured at $B = 9$ T and $T = 2$ K with $V_{bias} =$ 1mV. The interfacial resistance area product ($R_JS$) is 29.7 and 45.6 $\Omega$ μm$^2$ for spin-triplet MR device and control device, respectively. **d,** The MR results of control device (the same curve as in **c**). Clearly, the signal is much smaller and within the noise level. **e,** The comparison of MR in control (red symbols) and spin-triplet MR devices (blue symbols) with various $R_JS$ values. The blue and red stars correspond to the results of spin-triplet MR and control devices in **c**, respectively.

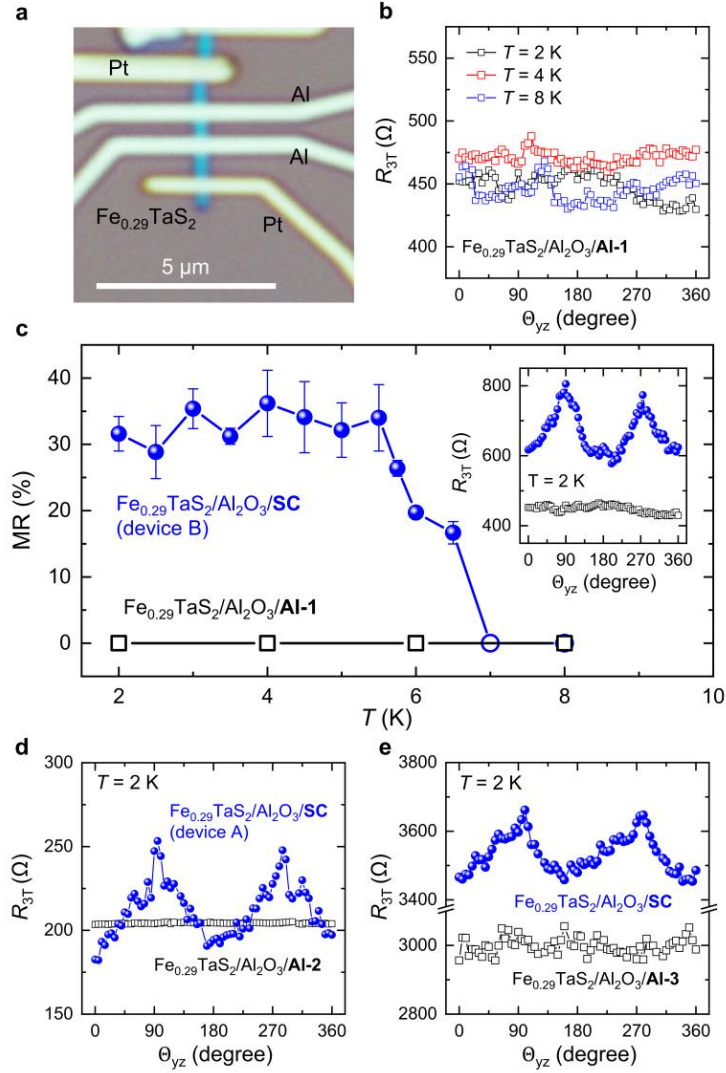

**Figure S9. Results of control devices of $Fe_{0.29}TaS_2/Al_2O_3$/normal metal (Al). a,** The optical images of a typical control device $Fe_{0.29}TaS_2/Al_2O_3$/Al, where the SC NbN electrode is replaced with ~ 50 nm Al. **b,** The MR results of the control device $Fe_{0.29}TaS_2/Al_2O_3$/Al-1 measured at $B =$ 9 T and $T =$ 2, 4, and 8 K, respectively. **c,** The comparison of the spin-triplet MR (blue symbols) and control device $Fe_{0.29}TaS_2/Al_2O_3$/Al (black symbols; $R_JS$: 59.4 Ω μm$^2$) as a function of temperature measured at $B =$ 9 T. Inset: The MR curves of control device and the spin-triplet MR device B with similar $R_JS$. **d, e,** The absence of MR signals on two other control devices $Fe_{0.29}TaS_2/Al_2O_3$/Al-2, and $Fe_{0.29}TaS_2/Al_2O_3$/Al-3 with $R_JS$ of 33.8 and 1626.2 Ω μm$^2$, respectively, measured at $B =$ 9 T and $T =$ 2 K. The blue symbols represent the MR curves on spin-triplet MR devices with similar $R_JS$.

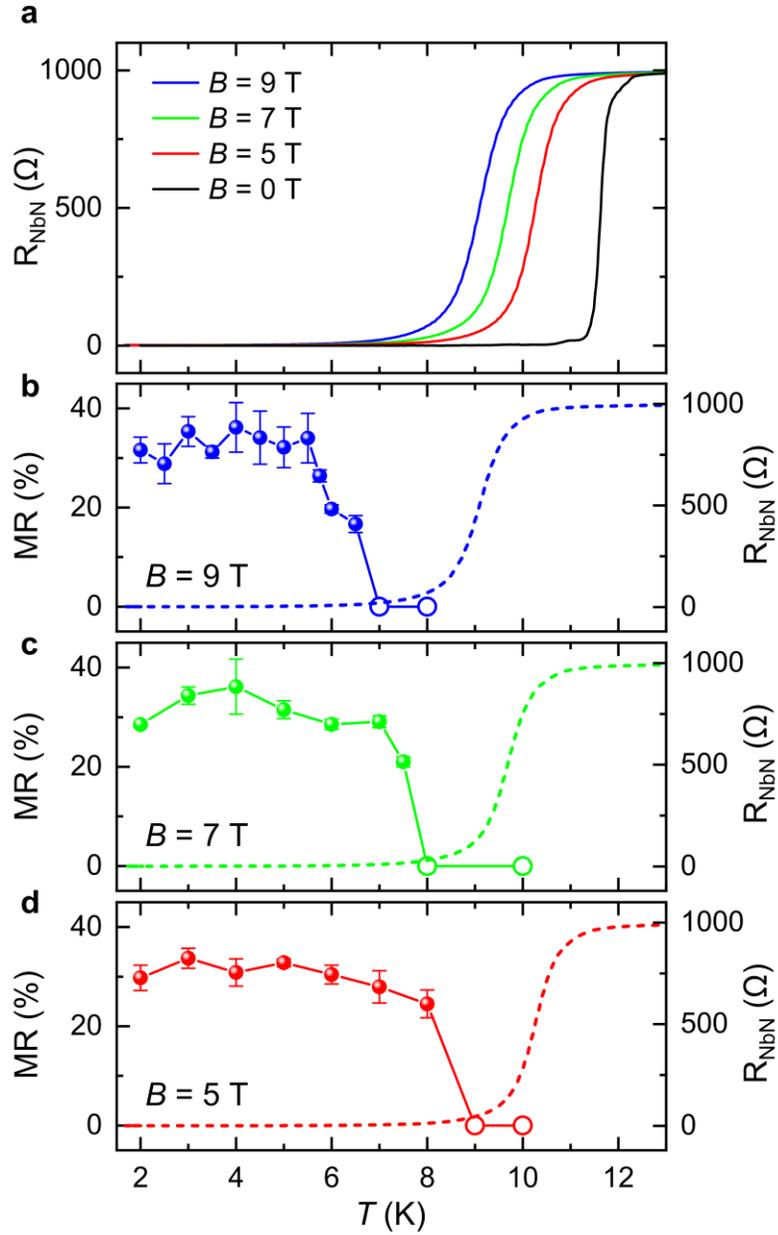

**Figure S10. The $T_{SC}$ of the NbN electrode and the correlation with the temperature-dependent MR. a,** The temperature dependence of the resistance measured on a typical NbN electrode under the perpendicular $B$ = 0, 5, 7, and 9 T, respectively. **b-d,** The temperature dependence of MR and the resistance of the NbN electrode (dashed lines) measured at $B$ = 9, 7, and 5 T, respectively.